# Strain Controlled Magnetic and Optical Properties of Monolayer 2H-TaSe$_2$

Sugata Chowdhury[1, 2], Jeffrey R. Simpson[1, 3], T. L. Einstein[4], Angela R. Hight Walker[1]

[1]Physical Measurement Laboratory, National Institute of Standards and Technology Gaithersburg, MD, 20899, USA; [2]Materials Measurement Laboratory, National Institute of Standards and Technology Gaithersburg, MD, 20899, USA; [3]Department of Physics, Astronomy, and Geosciences, Towson University, Towson, MD, 21252, USA; [4]Department of Physics and Condensed Matter Theory Center, University of Maryland, College Park, Maryland, 20742, USA

**Abstract:** First-principles calculations are used to probe the effects of mechanical strain on the magnetic and optical properties of monolayer (ML) 2H-TaSe$_2$. A complex dependence of these physical properties on strain results in unexpected behavior, such as ferromagnetism under uniaxial, in-plane, tensile strain and a lifting of the Raman-active $E'$ phonon degeneracy. While ferromagnetism is observed under compression along $\hat{x}$ and expansion along $\hat{y}$, no magnetic order occurs when interchanging the strain direction. The calculations show that the magnetic behavior of the system depends on the exchange within the 5d orbitals of the Ta atoms. The magnetic moment per Ta atom persists even when an additional compressive strain along the $\hat{z}$ axis is added to a biaxially-strained ML, which suggests stability of the magnetic order. Exploring the effects of this mechanical strain on the Raman-active phonon modes, we find that the $E''$ and $E'$ modes are red-shifted due to Ta-Se bond elongation, and that strain lifts the $E'$ mode degeneracy, except for the symmetrical biaxial tensile case. Our results demonstrate the possibility of tuning the properties of 2D-materials for nanoelectronic applications through strain.

**Introduction**

Two-dimensional (2D) layered materials have been the subject of intensive research due to their novel properties.[1-8] Among them graphene has been investigated most extensively. Electrical mobility and the Young's modulus of graphene are extremely high,[9] yet its band gap is zero. These properties have triggered a broad search for other novel 2D materials for electronic applications. In particular, monolayer (ML) transition metal dichalcogenides (TMDs) are attractive because of their graphene-like hexagonal structure[10] and versatile electronic properties.[11-14] Based on their electronic properties, ML-TMDs can be classified as insulators, semiconductors, semimetals, metals, and superconductors.[3] The electronic properties of ML-TMDs can be useful for fundamental and technological research in different fields. For example, ML-TMDs can be used as electrocatalysts for hydrogen production from water, electrodes in rechargeable batteries and photovoltaic cells, and field-effect transistors.[8, 15-20]

Experiments have shown that ML-TMDs can be fabricated by mechanical or solvent-based exfoliation methods,[21, 22] or produced by chemical vapor deposition.[23, 24] The properties of ML TMDs can be tuned by controlling chemical composition, functionalization, mechanical strain, and the application of external (electric and magnetic) fields.[5, 13, 14, 22, 25-31] Magnetic materials are an important component of low-dimensional nanoelectronic devices, but it is still rare to find a pristine, magnetic ML-TMD. Recently, different strategies have been proposed to introduce magnetism into TMDs, such as vacancy and impurity doping in bulk materials and edge effects in nanoribbons.[10, 28, 32-39] The successful fabrication of both TMD nanoribbons and TMDs with uniform defect density has proven complicated and remains an experimental challenge. Mechanical strain offers an attractive, alternative option because TMDs possess robust mechanical properties and can sustain about 11 % strain.[40] Several experimental and theoretical results have shown that the magnetic order in ML-TMDs[13, 14, 25, 26, 41-48] can be tuned by applying mechanical strain, and Wang *et al*. reported that optical properties can also be tuned with strain.[49] Previous theoretical work only considered symmetrical biaxial tensile[13, 14, 25, 26] or uniaxial tensile strain.[45] Thus, considering these previous studies, it is of great interest to determine how the metallic ground state of TMDs behaves under different directional strain.

Here we report magnetic and optical properties of pristine, ML-2H-TaSe$_2$ under uniaxial and biaxial tensile strain, as well as combinations of expansion and compression along their basal planes, within the framework of density functional theory (DFT). To understand the magnetic behavior, we consider three different spin structures, namely, nonmagnetic (NM), ferromagnetic (FM), and antiferromagnetic (AFM), which are shown in supplementary information (Fig. S1). We predict a complex relationship between mechanical strain and magnetic ordering in TaSe$_2$, wherein the magnetic order depends on the strength and the direction of the applied strain.

## Results

### Crystal Structure 2H-TaSe$_2$

TaSe$_2$ has four different phases, but for this work, we will consider only its 2H phase. The trigonal prismatic structure of 2H-TaSe$_2$ is illustrated in Fig S1a. Ground state calculations are performed by relaxing atomic position and lattice vectors of the bulk 2H-TaSe$_2$ (point group D$_{6h}$) structure. The optimized lattice constants ($a_{TaSe2}$ = 0.339 nm and c$_{TaSe2}$ = 1.22 nm) are in good agreement with previous computational studies[50, 51]. Relaxed structural parameters, such as bond length and intralayer distance, have been tabulated in supplementary information (Table S1). Starting with this relaxed bulk structure, we construct a ML supercell with 16-unit cells (4 x 4 x 1) and the resulting point group symmetry of D$_{3h}$.

### Applied Theoretical Strain

The effects of strain on the supercell are modeled by stretching and compressing along the $\hat{x}$- and/or $\hat{y}$-directions, as shown in Fig. S2. In this study, we define expansion as tensile strain and the combination of expansion and compression as shear strain.[52] The tensile strain was applied in three different ways as illustrated in Fig. S2: uniaxial expansion of ML-TaSe$_2$ in the (a) $x$-direction (XX$_+$), (b) $y$-direction (YY$_+$); and (c) homogeneous biaxial expansion in both $\hat{x}$- and/or $\hat{y}$-directions (XX$_+$YY$_+$). Also, we applied two types of the shear strain (d) expansion and compression in the monolayer along the $\hat{x}$- and/or $\hat{y}$-directions (XX$_+$YY$_-$), respectively, and (e) compression in the $x$-direction and expansion in $y$-direction (XX$_-$YY$_+$) with the same magnitude of strain [50]. The theoretically applied strain varies from 0 % to 12 % in 2 % increments. We limited the maximum strain to 12 % because an experiment showed that non-graphene TMDs rupture once strain reaches 11 %[40]. Lastly, we calculate the effect of applied compression along the $\hat{z}$-direction (ZZ$_-$) to examine the stability of magnetic order resulting from applied biaxial strain (XX$_+$YY$_+$). The evolution of our calculated lattice parameters and bond lengths (see Fig. S3) demonstrates smooth behavior with applied mechanical strain under various configurations and shows no sign of lattice rupture up to 12% strain. Using the couple cluster single double triple method (CCSD(T)) Iozzi et. al.,[53] found that 10 % mechanical strain can break a covalent. Additionally, Pan et. al.[13] theoretically showed that the magnetic properties of VTe$_2$ remained stable under 16 % mechanical strain.

### Magnetic Properties

To study the magnetic ordering under strain, we first compare the energies of the ferromagnetic (E$_{FM}$), antiferromagnetic (E$_{AFM}$) and nonmagnetic (E$_{NM}$), states. Figure 1(a) shows the energy differences between E$_{AFM}$ and E$_{FM}$ (ΔE$_{AFM-FM}$) and E$_{FM}$ and E$_{NM}$, (ΔE$_{FM-NM}$) as a function of mechanical strain. ML-TaSe$_2$ is a non-magnetic system in the absence of strain. The calculated energy differences indicate that ML-TaSe$_2$ becomes ferromagnetic when uniaxial or biaxial tensile strain of 6 % or larger are applied. The energy difference between AFM and FM ($\Delta E_{AFM-FM}$) is 24 meV with biaxial tensile strain of 6 %, and increases to 82 meV at 12 %. Surprisingly, under shear strain the magnetic ordering is different. When expansion along the *x* direction and compression along the *y* direction is applied [Fig S2 (d)], ML TaSe$_2$ remains non-magnetic for any amount of strain. However, it is ferromagnetic under 6 % strain, when the strain directions are reversed [Fig. S2(e)]. To find the origin of this complicated relationship between applied strain and magnetic ordering in ML-TaSe$_2$, we calculated the magnetic moment and charge transfer for each atom. Fig 1(b) displays dependence of the magnetic moment ($\mu_B$) of each atom in TaSe$_2$ as a function of applied strain. The contribution to the magnetic moment for each Ta atom increases when the strain exceeds 6 %, which is defined as the critical tensile strain.[13, 26, 45] At 6 % biaxial tensile strain, the magnetic moment is 0.33 $\mu_B$ and increases to 0.52 $\mu_B$ at 12 %. It increases with increasing biaxial and uniaxial tensile strain, as well as with shear strain XX$_-$YY$_+$ (compression along the $\hat{x}$-axis and expansion along the $\hat{y}$-axis). Finally, it is negligible when tensile strain along x and compression along y are applied. The total magnetic moment arises mainly from the Ta atoms, with a negligible contribution from the Se atoms.

The calculated charge transfer between Ta and Se atoms is shown in Fig. 1(c). We use two different methods for charge transfer analysis: (1) Lowdin[54, 55] and (2) Bader charge analysis.[56] Here we presented only Bader charge analysis. At 6 % biaxial tensile strain, the Ta atom loses about 0.19 electrons, whereas the Se atom gains only 0.05 electrons, resulting in an increased electron density on the Ta atoms. This implies that the magnetic behavior of the system depends on the exchange between spins in the 5d orbitals of the Ta atoms. The exchange process among the 5d orbitals of the Ta atoms increases with increasing strain, rising to 0.3 electrons at 12 % strain. This exchange is the same under any type of applied uniaxial strain and one type of shear strain XX$_-$YY$_+$. However, the system is metallic, and the exchange process among the 5d orbitals of the Ta atoms is negligible when the other XX$_+$YY$_-$ shear strain is applied.

We found that, the bond length between Ta-Se atoms increases with exchange between the spins in the 5d orbitals of Ta atoms and decreases only when expansion along $\hat{x}$-axis and compression along $\hat{y}$-axis (XX$_+$YY$_-$) is applied (see Table S2). We conclude that the spin exchange process and bond lengthening depend on the type and direction of applied strain.

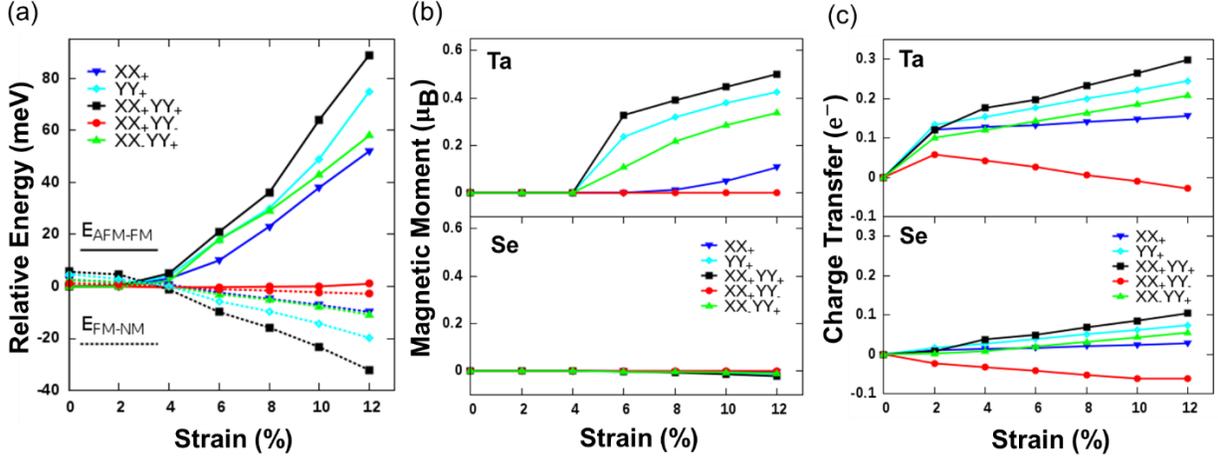

**Fig. 1.** (a) The calculated energy differences per unit cell between different magnetic orders of ML-TaSe$_2$ as a function of applied strain. The energy difference between different antiferromagnetic (AFM) and ferromagnetic (FM), $\Delta E_{AFM-FM} = E_{AFM} - E_{FM}$; is plotted as the top set of curves; and the difference between non-magnetic states (NM) and ferromagnetic states (FM), $\Delta E_{FM-NM} = E_{FM} - E_{NM}$] is the bottom set. The magnetic moment (b) and the electron charge transfer (c) per Ta atom (top) and per Se atom (bottom) of ML-TaSe$_2$ structures using different strain approaches.

An individual Ta atom is magnetic, but covalent interactions and strong hybridization, remove the Ta atom's magnetic moment. Therefore, our goal is to understand the nature of spins in the 5d orbitals of individual Ta atoms under different types and directions of applied strain. To achieve this, we compare the partial density of states (PDOS) of unstrained and strained ML-TaSe$_2$. In Fig. 2, the red and blue regions represent the spin-up and spin-down components, respectively. In Fig. 2(a) and (b) we show the PDOS of the 5d orbital of Ta and the 4p orbital of Se atom without strain (0 %) and under 6 % biaxial tensile strain, respectively. At the Fermi level, the PDOS decreases in the 4p states of the Se with 6 % strain [Fig 2b] and these states no longer take part in the hybridization process with the Ta atoms. On the other hand, the spin degeneracy is broken for the 5d orbitals of the Ta atom as highlighted with arrows [Fig 2b], and magnetic order is induced in the system. At this Fermi level, the character of the 5d orbitals of the Ta atom is comprised of the $d_{x^2-y^2}$ and $d_{xy}$ orbitals, which lie in the basal plane. The spin-splitting exchange between these two orbitals is enhanced as the strain is increased, and is directly proportional to the bond length between the Ta and Se atoms. In contrast, under XX$_+$YY$_-$ strain shown in Fig. 2(c), the Se 4p PDOS of increases and the Ta 5d PDOS decreases. This result suggests that the spin polarization increases as the hybridization process between the Ta and Se atoms decreases. Finally, with biaxial compressive strain ML-TaSe$_2$ is found to be a NM system due stronger hybridization.

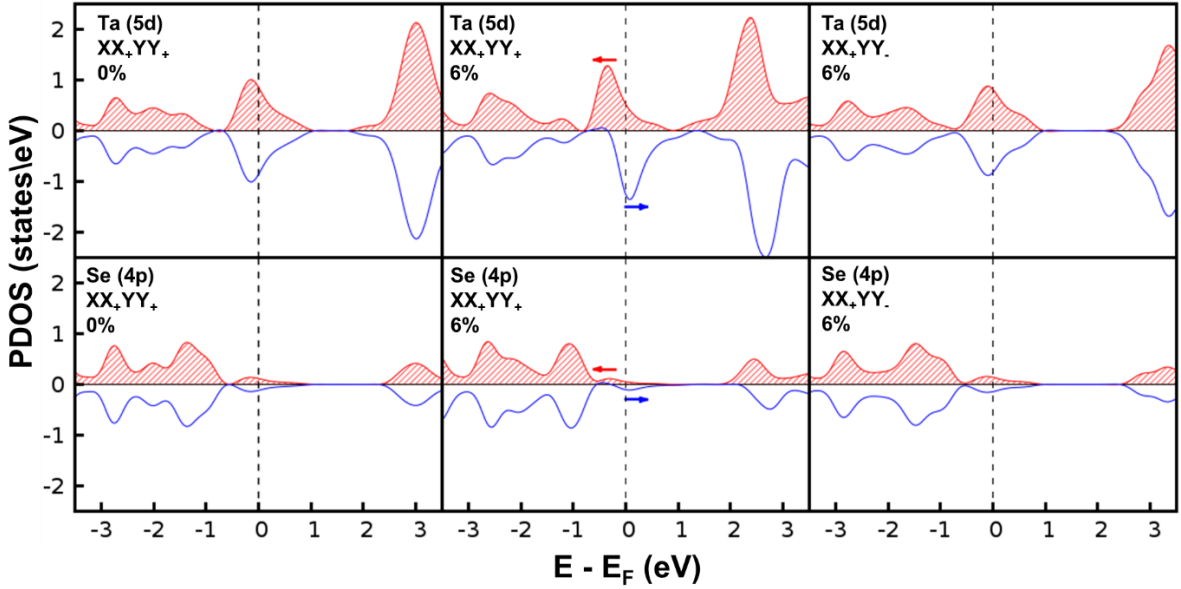

**Fig. 2**. Spin-polarized, partial density of states (PDOS) of Ta (top) and Se (bottom) corresponding to (a) 0 %, (b) 6 % symmetrical biaxial tensile strain, and (c) 6 % expansion in the $\hat{x}$-direction and compression in the $\hat{y}$-direction (XX$_+$YY$_-$). Here red represents the spin-up and blue represents the spin-down. The Fermi level is indicated by the dashed vertical black line and the horizontal arrows highlight the spin splitting near the Fermi level, which is only non-zero in (b) for 6 % biaxial strain.

Considering practical applications in nanoelectronic devices, we also determine the effect of magnetic ordering with additional strain is applied along the z-axis. We consider two different situations along the z-axis: (1) application of compressive strain (ZZ$_-$) when ML-TaSe$_2$ is under no symmetrical biaxial tensile strain, and (2) compression along Z with biaxial tensile strain at 6 %. The magnetic moment ($\mu_B$) of each Ta atom under strain along $\hat{z}$-direction at 6 % symmetrical tensile strain (XX$_+$YY$_+$) of ML-TaSe$_2$ is shown in Fig 3. The magnetic moment per Ta atom is 0.09 $\mu_B$ when compressive strain is applied along the z-axis without additional strain in the basal plane, and the intralayer distance between Se-Se is 3.24 Å. Under symmetrical biaxial strain (6 %), the Ta magnetic moment starts at 0.33 $\mu_B$ and reduces to 0.24 $\mu_B$ with applied strain of 5 % along the z-axis. If we apply compression along the z-axis of greater than 5 %, the magnetic order is destroyed. This implies that the magnetic order is destroyed due to strong covalent bonds between Ta-Se.

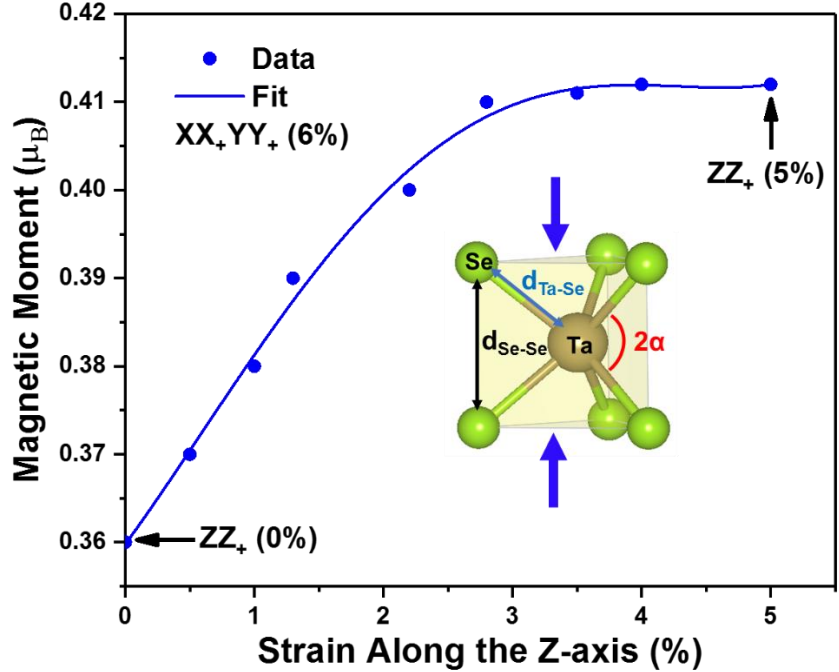

**Fig. 3**. Magnetic moment ($\mu_B$) of each Ta atom under strain along the direction ($\hat{z}$) perpendicular to the XY plane at 6 % symmetrical tensile strain (XX$_+$YY$_+$) of ML-TaSe$_2$. Inset: the trigonal prismatic coordination of 2H-TaSe$_2$ and applied strain direction along ZZ-axis. These results suggest that magnetism is sustained even with applied strain along the Z-axis.

Also, the calculated Curie temperature ($T_C$) was used to probe the stability of the magnetic order at high temperature. While it is long known that mean field theory can drastically overestimate $T_c$ in two-dimensional systems,[57, 58] this effect may be less problematic for long-range interactions (the Heisenberg model) and has been routinely used in analyzing nanoparticles[26] and surface of magnetic systems.[59] Furthermore, even if mean field estimates of $T_c$ are too large, the trend should be meaningful. Thus, we estimate $T_C$ from $k_B T_C = (2/3)\Delta E_{AFM-FM}$. Our enhanced $T_C$ is approximately 287 K at 8 % and nearly twice that at 12 % biaxial XY tensile strain. This finding suggests that ferromagnetic behavior of strained TaSe$_2$ monolayer is stable at high temperature and it could be useful for nanoelectronic devices operating in such temperature range.

**Optical Properties – Raman Spectroscopy**

In this section, we discuss the optical properties of ML 2H-TaSe$_2$, specifically the evolution of the Raman-active phonon modes due to applied mechanical strain. Raman spectroscopy provides detailed information about the lattice structure of materials. To probe the effect of strain on the Raman phonons, DFT predictions are compared with experimental results for bulk 2H-TaSe$_2$ ($D_{6h}$). At the Γ point, bulk 2H-TaSe$_2$ has twelve phonon modes, represented by Γ ($D_{6h}$) = $A_{1g}$ + 2$A_{2u}$ + $B_{1u}$ + 2$B_{2g}$ + $E_{1g}$ + 2$E_{1u}$ + $E_{2u}$ + 2$E_{2g}$ although only four are Raman active: $A_{1g}$, $E_{1g}$, and 2 $E_{2g}$ as shown in Fig 4a. The Raman spectrum obtained at room temperature is shown in Fig. 4(c) with data as blue dots and peak fit as a solid blue line. Our experimental results, in agreement with previous studies,[60] show three main peak features at 140 cm$^{-1}$, 209 cm$^{-1}$ and 237 cm$^{-1}$. The DFT results agree extremely well with our experiments giving predicted values for $E_{1g}$, $E_{2g}$, and $A_{1g}$ modes at 139 cm$^{-1}$, 213 cm$^{-1}$, and 243 cm$^{-1}$.

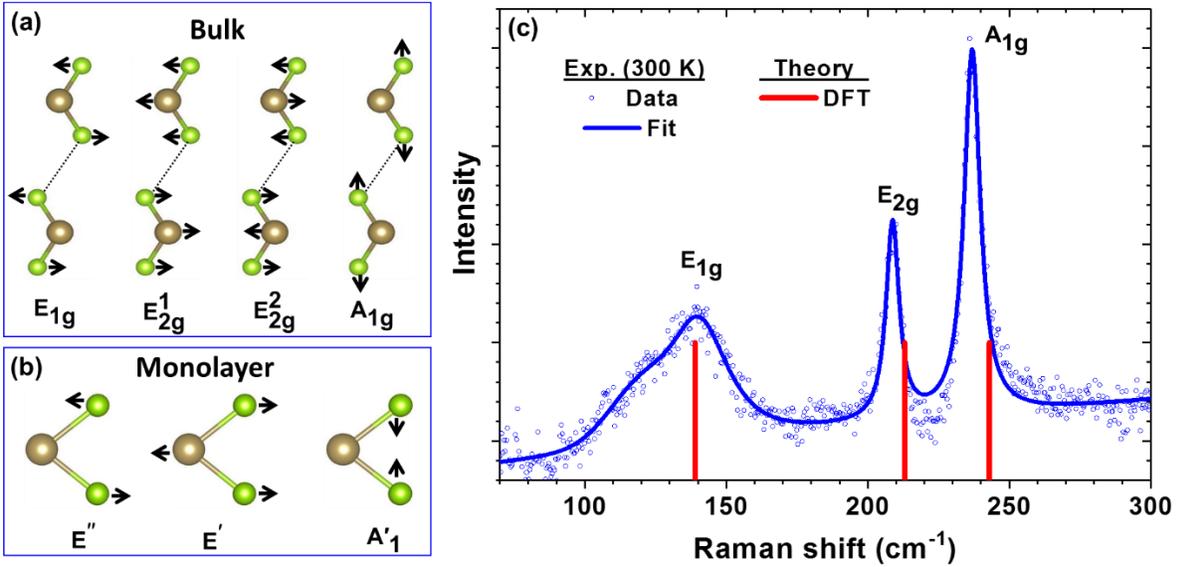

**Fig. 4.** Symmetry allowed Raman active modes of unstrained 2H-TaSe$_2$ for (a) bulk and (b) monolayer. (c) Comparison between Raman spectra of unstrained bulk 2H-TaSe$_2$ measured experimentally at 300 K (blue circles and blue fit line) and calculated prediction using DFT (red line).

Next, we examine the Γ-point phonon modes of ML-TaSe$_2$, whose optical properties are different from those of bulk 2H-TaSe$_2$ due to differences in point group and dimensionality. The point group symmetry of ML-TaSe$_2$ is $D_{3h}$, with six phonon modes: $A'_1$, $A'_2$, $E'$, $A''_1$, $A''_2$, and $E''$, only three of which are Raman-active: $E''$, $E'$, and $A'_1$. In Fig. 4(b), the Raman active modes of ML-TaSe$_2$ are shown, which correspond to the characteristic $A_{1g}$, $E_{1g}$, and $E_{2g}$ modes of bulk 2H-TaSe$_2$. We also compared our predictions with other theoretical predictions[51] and experimental results.[60] We predict Raman peaks at 138 cm$^{-1}$ ($E''$), 210 cm$^{-1}$ ($E'$), and 237 cm$^{-1}$ ($A''_1$), which are in good agreement with previous reports.[51, 60]

Fig. 5 shows the calculated frequency shifts of Raman-active phonon modes at the Γ-point as a function of strain from 0 % to 12 % for the indicated strain configurations. Our analysis reveals that $E''$ is the most sensitive mode and shifts to lower frequency (red shifts) under all strain conditions. The magnitude of the $E''$ red shift depends on the type of applied strain. The red shift of $E''$ is a maximum (~40 cm$^{-1}$) under symmetrical biaxial strain (12 %) and a minimum (~10 cm$^{-1}$) under XX$_+$YY$_-$. The $A'_1$ peak blue shifts, reaching maximum (~10 cm$^{-1}$) under symmetrical biaxial strain (12%). The most interesting Raman peak is $E'$, which involves opposite vibrations of the two Se atoms with respect to the Ta atom in the basal plane [Fig. 4(b)]. The degeneracy of $E'$ is lifted upon application of uniaxial or pure shear strain. This doubly-degenerate optical phonon mode ($E'$) further splits into two singlet sub-bands, named $E'^+$ and $E'^-$ for higher and lower energy, respectively. An obvious red shift occurs, shifting by approximately 40 cm$^{-1}$ under 12 % symmetrical biaxial strain.

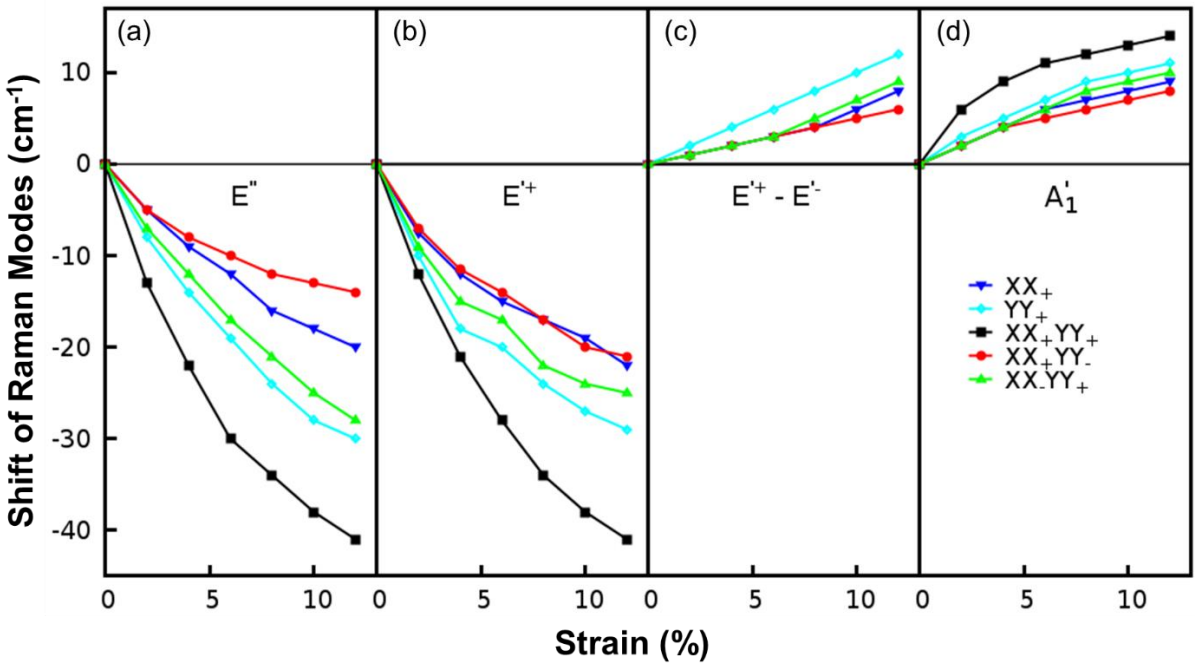

**Fig 5.** The shift of the Raman modes of ML-TaSe$_2$. Panel (a) and (b) represent in-plane shearing modes $E''$ and $E'^+$, respectively and panel (d) represents the out-of-plane mode $A'_1$. (c) Peak frequency difference between $E'^+$ and $E'^-$. $E'$ is no longer degenerate with each $E'$ split into two modes: $E'^+$ and $E'^-$. These modes are present for all types of strain, except the symmetrical biaxial strain XX$_+$YY$_+$ (black squares).

To understand the mechanical strain effect on the optical properties of ML-TaSe$_2$, we calculated the phonon dispersion curves (Fig. 6), both without strain (0%) and under 6% strain in the indicated configurations. Changes in all Raman modes under strain are shown in Fig. S8-12. To reduce the computational time, we considered unit cells that contain only three atoms for all ML TaSe$_2$ and produce nine phonon bands. Out of the nine, the three acoustic bands (red in Fig. 6) are separated from the six optical bands (blue in Fig. 6) by a 20 cm$^{-1}$ indirect phonon frequency gap. Analyzing the phonon dispersion curve, we noticed that there are no negative phonon frequencies in the acoustic branch in the unstrained (0 %) or 6% symmetrical biaxial strain (Fig. 6(a)). Also, there is no splitting within the optical branches. Interestingly, all other kinds of strain produce negative phonon frequencies for the acoustic branch along the ΓM, MK, and KΓ directions and cause splitting in the optical branches. Negative frequencies for acoustic modes imply instability to a distorted structure. This negative phonon dispersion is seen in the charge density wave phase (CDW) of TaSe$_2$ at low temperature (122 K).[61] The MK direction has the largest effect and has not been reported previously. Due to mechanical strain, the monolayer structure is distorted and elongated. These results can be understood by noting that, compared to symmetrical biaxial strain, uniaxial tensile strain and pure shear strain break symmetry in these materials, leading to a new phase.

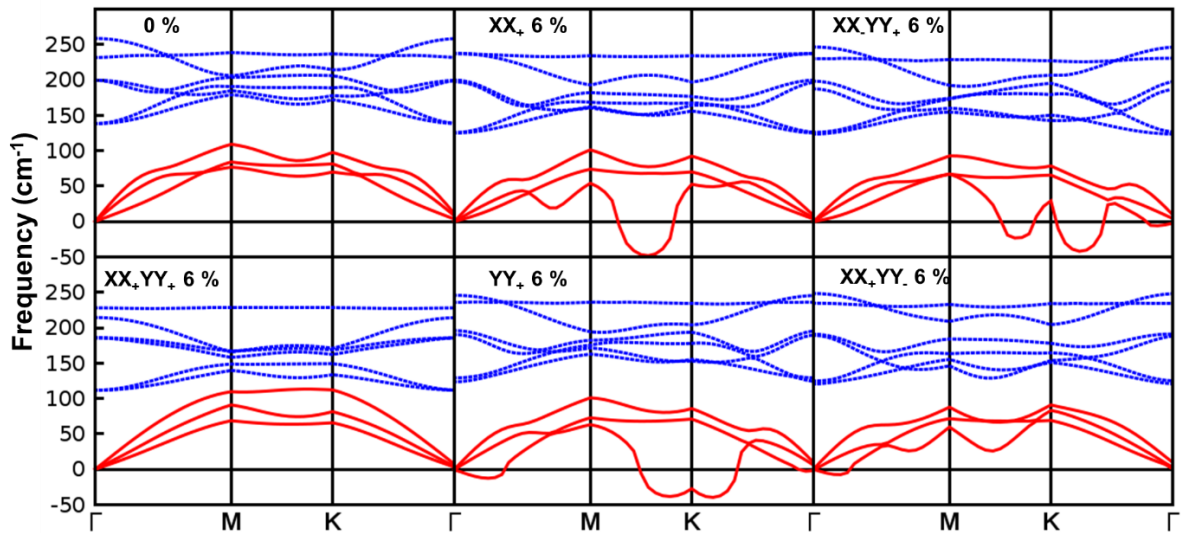

**Fig 6**. Predicted phonon dispersion curves of ML-TaSe$_2$ in **k**-space as a function of different types of in-plane strain compared to the unstrained system (upper left panel). Blue and red dispersion curves represent the optical and acoustic modes, respectively. The phonon dispersion curves are substantially changed compared to the unstrained and symmetrical biaxial strain case (lower left panel). Strain lifts the degeneracy of the $E'$ mode at 184 cm$^{-1}$ frequency in YY$_+$ strain (upper middle and upper right panels). Negative frequencies for acoustic modes imply instability to a distorted structure.

**Conclusion**

In summary, we studied the magnetic and optical properties of monolayer TaSe$_2$ under mechanical strain using spin-polarized DFT. Pristine, ML 2H-TaSe$_2$ is nonmagnetic but becomes ferromagnetic with applied basal strain beginning at approximately 6 %, except for pure shear strain, *e.g.*, simultaneous symmetrical tensile strain along the x-axis and compressive strain along the y-axis. The magnetic order of metallic TaSe$_2$ depends on the direction of the applied strain. The calculated Curie temperature of strained TaSe$_2$ implies that the ferromagnetic ML is suitable for spintronics applications at room temperature. Most importantly, we demonstrate the stability of the magnetic order under additional pressure along the z-axis, a robustness which is desirable for practical applications of TMDs in nanoelectronic devices. The experimental Raman spectrum of bulk 2H-TaSe$_2$ and our DFT predicted phonon frequencies are in excellent agreement. Raman calculations of strained ML-TaSe$_2$ show significant redshift of $E''$ and blueshift for $A_1'$. We also find that the doubly-degenerate $E'$ mode splits into two components: $E'^+$ and $E'^-$. This splitting and shifting of the modes depends on the direction and magnitude of the applied strain. The phonon dispersion evaluation with respect to mechanical strain is intriguing, particularly with similarities to other CDW materials. Our work sheds light on TMD materials and their possible applications in strained devices.

**Methods**

**Computational:** Calculations were carried out using density-functional theory (DFT)[62, 63] as implemented in PWSCF code,[64] using the projector augmented wave (PAW) method.[65] Within the local-density approximations (LDA), we employ Perdew-Zunger (PZ) exchange and correlation functionals.[63] The kinetic energy cutoff of the plane-wave expansion is taken as 520 eV. All the geometric structures are fully relaxed until the force on each atom is less than 0.002 eV/Å, and the energy-convergence criterion is 1x10$^{-6}$eV. For the electronic-structure calculations, a 32x32x1 **k**-point grid is used. To estimate the charge transfer we have use the Bader charge analysis code.[66] To investigate the spin structure, we construct a 4x4x1 supercell and use a sufficiently large vacuum (20 Å) in the vertical direction to render negligible any interaction between neighboring supercells.

**Raman Measurements**

Raman spectrum for bulk TaSe2 excited with a HeNe laser at 632.8 nm (sample power < 1mW) was measured at room temperature using a triple-grating Raman spectrometer (Horiba JY T64000 with a 50X objective, N.A. 0.82) coupled to a liquid-nitrogen cooled CCD detector. All spectra were taken in the 180° backscattering configuration. Bulk TaSe$_2$ sample was freshly exfoliated.

The purpose of identifying the computer software and experimental setup in this article is to specify the computational procedure and Raman measurement. Such identification does not imply recommendation or endorsement by the National Institute of Standards and Technology.

## Acknowledgement
We thank Dr. Kamal Choudhary and Francesca Tavazza at NIST for helpful discussions.

## Contributions

S.C. performed all DFT calculations and worked on data analysis, verification and writing the manuscript. A.R.H.W and J.S. performed the experiments, provided scientific discussions and assisted in writing the manuscript. T.L.E. assisted in writing the manuscript and provided scientific discussions. All the DFT calculations were done at using NIST supercomputing center.

## Conflicting Interests

The authors declare that they have no financial or non-financial conflicting interests.